\documentclass[twocolumn]{aastex631}

\usepackage{hyperref} %for references
\usepackage{graphicx} %for adding plots
\usepackage{epstopdf}
\usepackage{amsmath} % Provides advanced math functionalities
\usepackage{orcidlink} % For \orcidlink command

\begin{document}

\title{Solar Periodic Companion and Random Stellar Flybys: 
Dynamical Perturbations of Highly Eccentric Comets in the Oort Cloud}

\author[0009-0005-4168-2858]{Shahed~Shayan~Arani \orcidlink{0009-0005-4168-2858}}
\email{sshayanarani@ucsd.edu}
\affiliation{Department of Physics, University of California, San Diego, La Jolla, CA 92093, USA}
\affiliation{Center for Dark Cosmology and Gravitation, University of California, San Diego, La Jolla, CA 92093-0319}

\begin{abstract}
The \textit{\textit{Gaia}} space telescope has transformed our understanding of random stellar encounters with the Solar System. This study examines how such perturbations influence the most eccentric comets in the Oort Cloud (OC), a thermalized reservoir of $\sim10^{12}$ icy bodies extending from $10^{4}$ to $10^{5}$~AU. Recent \textit{\textit{Gaia}}-based analyses indicate $\sim20$ stellar passages within 1~pc of the Sun per Myr. Using analytical estimates and direct $N$-body simulations, we quantify how these encounters modify highly eccentric orbits: individual flybys enhance comet-shower rates by factors of $\sim2$, producing a cumulative increase of $\sim40$ over a Myr. In parallel, we perform a full dynamical search through all stars with six-dimensional phase-space data in \textit{\textit{Gaia}}~DR3 and identify a compelling candidate for a long-period stellar companion to the Sun. This star exhibits recurrent $\sim45$~Myr perihelion passages and, in simulations, can elevate comet-shower activity by an order of magnitude. Because a perturber of this kind could also be a dispersed solar sibling, the companion hypothesis links the dynamical structure of the OC directly to the long-standing problem of recovering the Sun's birth-cluster relatives. The chemical-abundance pattern of such a star therefore becomes a critical observational test. Together, these results clarify how both stochastic encounters and potential long-period companions shape the dynamical evolution and observable output of the outermost OC, while highlighting the possibility that the OC retains a dynamical memory of the Sun’s earliest stellar environment.

\end{abstract}


\section{Introduction}

The dynamical evolution of the Oort Cloud (OC) is governed by a combination of Galactic tides, stellar flybys, and long-term perturbations arising from the Sun’s motion in the Milky Way. While the OC is too distant to be observed directly, its structure and history are imprinted in the flux and orbital distribution of long-period comets entering the inner Solar System. Consequently, understanding the mechanisms that perturb the most eccentric comets is crucial for connecting the present-day comet population to its dynamical origins. Over the past decade, the \textit{\textit{Gaia}} mission has fundamentally transformed this problem by providing precise six-dimensional phase-space measurements for tens of millions of stars, enabling a statistically complete characterization of stellar encounters and a direct dynamical search for stars with long-term periodic relationships to the Sun.

Stellar flybys have long been recognized as a dominant heating mechanism for the OC, reshaping orbital energies and diffusing comets into the planetary region. Early work based on the \textit{Hipparcos} sample of $\sim10^{5}$ stars estimated a rate of roughly $11.6\pm1.3~\mathrm{Myr^{-1}}$ for encounters within 1~pc of the Sun \citep{Garcia_Sanchez_2001}. With the much deeper \textit{\textit{Gaia}}~DR2 catalog, the encounter rate nearly doubled to $19.7\pm2.2~\mathrm{Myr^{-1}}$ \citep{Bailer_Jones_2018}, confirming that close stellar passages are a regular feature of the Sun’s Galactic environment. These perturbers typically have velocities near $50~\mathrm{km\,s^{-1}}$ and masses close to that of the Sun, and their cumulative effect drives the OC toward a thermalized state \citep{Hills_1981,Heisler_1986,Fouchard_2017}. Analytical treatments show that the injection efficiency of comets into the inner Solar System depends strongly on encounter distance and velocity, while numerical simulations have clarified how these impulses alter the perihelia and angular momenta of the most eccentric comets. The \textit{\textit{Gaia}} dataset now allows these theoretical expectations to be anchored to measured encounter statistics.

Alongside the classical problem of random stellar perturbations, a parallel line of inquiry concerns the possibility of a long-period stellar companion to the Sun. Although previous searches have yielded null results, the idea persists due to several independent motivations. Scenarios involving the Sun’s birth cluster \citep{portegieszwart2009}, proposed outer Solar System bodies such as Planet~9 \citep{Batygin_2019,Pfalzner_2020}, and hypothesized periodicity in extinction events \citep{Whitmire_1984,Davis_1984} have all prompted searches for stars that could share a quasi-periodic orbital history with the Sun. Identifying such a companion---even if weakly bound or currently unbound---would carry implications for the dynamical history of the Solar System and the present-day structure of the OC.

Previous searches have relied on two complementary methods. Chemical tagging uses high-resolution spectroscopy to identify stars with solar-like elemental abundances \citep{Adibekyan2018}, while dynamical analyses use astrometry and kinematics to identify stars whose orbits are consistent with a shared origin or recurrent close approaches, as demonstrated in HARPS, AMBRE, APOGEE, GALAH, and \textit{\textit{Gaia}}-based studies \citep{Batista2014,Webb2019}. Although no definitive sibling or companion has yet been confirmed, chemo-dynamical tagging has proven powerful in other settings, such as the identification of extra-tidal halo stars escaping globular clusters \citep{Grondin2023,Grondin2024}. The precision of \textit{\textit{Gaia}}~DR3, with complete phase-space parameters for 33 million stars, now permits a comprehensive dynamical search for solar companions on long-period orbits.

In this work, we combine these two perspectives by examining the perturbative effects of individual stellar encounters on the most eccentric comets in the OC and by performing a systematic search for periodic stellar companions using the full phase-space precision of \textit{\textit{Gaia}}~DR3. First, we use direct $N$-body simulations to quantify how a single solar-mass stellar flyby alters the perihelion distribution, angular momentum structure, and comet-shower flux of a high-eccentricity OC ensemble. These simulations isolate the short-term dynamical response of comets and allow us to connect analytical scaling relations to concrete orbital outcomes. Second, we integrate the orbits of all $\sim33$ million \textit{\textit{Gaia}}~DR3 stars, along with the Sun, forward for 500~Myr in a realistic Galactic potential that includes axisymmetric components as well as the bar and spiral arms. From these trajectories we identify stars that exhibit repeated low-velocity encounters with the Solar System, evaluate their dynamical effects on a modeled OC, and assess their potential as long-period perturbers or solar sibling candidates.

The remainder of this paper is structured as follows. Section~2 develops the encounter-driven framework for Oort Cloud dynamics and presents the suite of $N$-body simulations used to quantify the response of highly eccentric comets to stellar flybys. Section~3 introduces the time-dependent Galactic potential adopted in this study, details the procedure used to search for long-period stellar companions in the \textit{Gaia}~DR3 dataset, and evaluates the dynamical signatures of the candidate star identified by this analysis, including its possible impact on the Oort Cloud. Section~4 summarizes the main results and discusses their implications for comet-shower statistics, the Sun’s birth environment, and avenues for future chemo-dynamical follow-up.

\begin{figure*}[t]
    \centering
    \includegraphics[width=\textwidth]{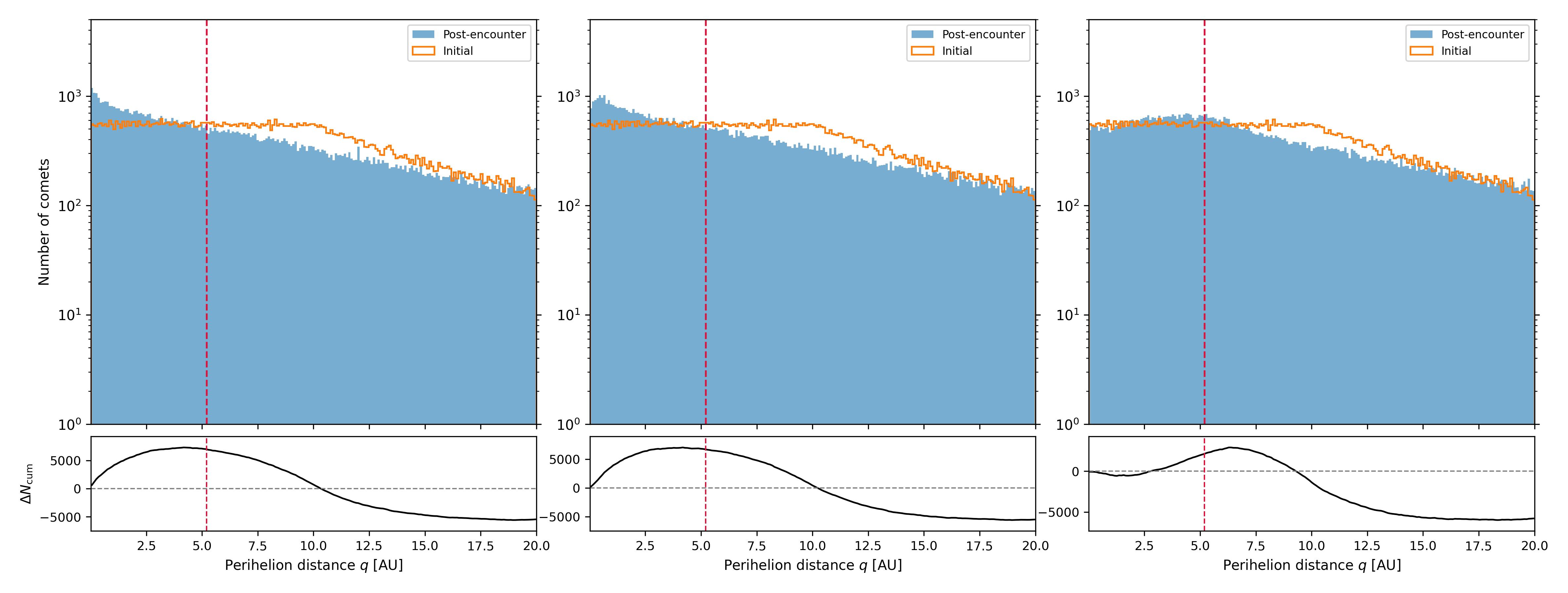}
    \caption{The perihelia of the comets before and after the encounter. The influx of comets entering the region with perihelia $<5$~AU is evident because the simulation isolates the most eccentric comets in the Oort Cloud. The left, middle, and right panels correspond to encounter distances of $2\times10^{5}$, $2\times10^{4}$, and $2\times10^{3}$~AU, respectively. As the perturber penetrates the inner bound of the OC, the total number of comets ejected within the Jupiter's orbit decreases compared to the other two instances, as expected from the considerations with regards to the angular momentum.}
    \label{peri_se}
\end{figure*}

\section{Stellar Encounters}

The \textit{Hipparcos} mission provided the six-dimensional phase-space parameters of roughly $10^{5}$ stars in the solar neighborhood, forming the first systematic dataset for studying stellar kinematics near the Sun. \textit{\textit{Gaia}}~DR3 expanded this dataset by more than two orders of magnitude, including approximately $3.3\times10^{7}$ stars with complete phase-space parameters observation. This dramatic increase in the available phase-space coverage has allowed a far more detailed analysis of the frequency and kinematics of stellar encounters with the Solar System. The rate of stars passing within 1~pc of the Sun, derived from \textit{Hipparcos} data, was estimated as $11.6 \pm 1.3$~Myr$^{-1}$ \citep{Garcia_Sanchez_2001}. Subsequent analysis using \textit{\textit{Gaia}}~DR2 nearly doubled this value to $19.7 \pm 2.2$~Myr$^{-1}$ \citep{Bailer_Jones_2018}. The velocity distribution of these encounters is approximately centered at 50~km~s$^{-1}$, with a spread of $\sim 20$~km~s$^{-1}$, and the characteristic masses of the perturbers are close to one solar mass. 

The dynamical effects of random stellar perturbations on OC comets have been investigated extensively in the literature \citep[e.g.,][]{Hills_1981,Heisler_1986,Fouchard_2017}, and have shed light on how repeated encounters gradually randomize the phase-space distribution of long-period comets. The analytical framework developed by J.~G.~Hills \citep{Hills_1981} remains particularly instructive and provides a theoretical picture of how stellar impulses diffuse cometary velocities over time. The essential definitions and relations from that work are summarized here to set the context for the numerical analysis that follows.

Hills introduced two key concepts—the \textit{loss cone} and the \textit{smear cone}. The loss cone refers to the Sun-centered bundle of velocity vectors in velocity space corresponding to cometary orbits whose perihelia lie within the planetary region. If a comet’s velocity vector falls within this cone, its orbit penetrates the inner Solar System and may lead to an observable comet shower. The smear cone, in contrast, represents the angular domain in velocity space through which a passing star randomizes cometary velocities during an encounter. The cumulative dynamical evolution of the OC can thus be viewed as the diffusion of cometary velocity vectors through successive smear cones that continually repopulate the loss cone.

For an individual stellar encounter, the impulse approximation provides a first-order estimate of the velocity perturbation imparted to each comet. Assuming that the stellar trajectory relative to the Sun can be treated as a straight line—a valid approximation for encounter velocities much larger than the local orbital velocities—the instantaneous impulse imparted to an object $i$ (either the Sun or a comet) is
\begin{equation}
    \Delta V_{i} = \frac{2GM_{\star}}{V_{\star} D_{i}},
\end{equation}
where $M_{\star}$ and $V_{\star}$ denote the mass and relative velocity of the passing star, respectively, and $D_{i}$ is the distance of closest approach. For the Sun, $D_{i}=D_{\odot}$ and $\Delta V_{i}=\Delta V_{\odot}$, whereas for a comet, $D_{i}=D_{c}$ and $\Delta V_{i}=\Delta V_{c}$. Because the two velocity impulses are approximately parallel, the differential velocity imparted to the comet relative to the Sun is given by
\begin{equation}
    \Delta V = \left| \Delta V_{c} - \Delta V_{\odot} \right| = \frac{2GM_{\star}}{V_{\star}}
    \left|\frac{D_{\odot}-D_{c}}{D_{c}D_{\odot}}\right|.
\end{equation}
Averaging this expression over all comets at distance $r$ from the Sun and adopting the straight-line approximation yields
$\langle |D_{\odot}-D_{c}| \rangle \approx r$ and $\langle D_{\odot} D_{c} \rangle \approx D_{\odot}^{2}$.
The time-averaged separation of two objects in Keplerian orbits with semi-major axis $a$ and eccentricity $e$ is
$\langle r \rangle = a(1+e^{2}/2)$, which approaches $\langle r \rangle \simeq 1.5a$ for nearly parabolic orbits ($e \to 1$).
With these approximations, the mean impulse becomes
\begin{equation}
    \langle \Delta V \rangle \simeq \frac{3a G M_{\star}}{V_{\star} D_{\odot}^{2}}.
\end{equation}
From conservation of energy, the characteristic orbital velocity of a comet with semi-major axis $a$ at $r=1.5a$ is
$V_{c} = (G M_{\odot} / 3a)^{1/2}$.

For weak perturbations where $\Delta V / V_{c} \ll 1$, the ratio of the angular area of the smear cone to that of the loss cone—representing the fractional diffusion of comets into the planetary region per encounter—is given by
\begin{equation}
    \frac{F_{s}}{F_{q}} = \frac{27}{8}
    \left(\frac{M_{\star}}{M_{\odot}}\right)^{2}
    \left(\frac{a}{D_{\odot}}\right)^{4}
    \left(\frac{G M_{\odot}}{a V_{\star}^{2}}\right).
\end{equation}
This equation encapsulates the essential scaling of encounter-driven injection: the rate of comet showers depends steeply on the fourth power of the stellar closest-approach distance ($D_{\odot}^{-4}$) and quadratically on both the stellar mass and the inverse encounter velocity. As a consequence, a small number of slow, massive, and close encounters dominate the dynamical evolution of the OC, even though the majority of encounters are distant and dynamically negligible.

Repeated random stellar perturbations gradually thermalize the orbital energy distribution of the OC comets. When the ensemble reaches statistical equilibrium, the fraction of orbits with eccentricities between $e$ and unity follows $F_{e}=1-e^{2}$, independent of semi-major axis. Given the relationship between perihelion distance $q$, semi-major axis $a$, and eccentricity $e$ through $q=a(1-e)$, the fraction of comets with semi-major axis $a$ that pass within a distance $q$ of the Sun is
\begin{equation}
    F_{q} = \frac{2q}{a}\left(1-\frac{q}{2a}\right),
\end{equation}
which quantifies the expected population of comets capable of penetrating the inner Solar System for a given energy distribution.

Complementary to the analytical approach of Hills, \citet{Fouchard_2017} provided an empirical fit based on Monte Carlo simulations involving $10^{6}$ test comets to estimate the number $N_{\star}$ of comets injected into the planetary region by a single stellar passage. Their results can be expressed as
\begin{equation}
    N_{\star} = 16.23
    \left(\frac{M_{\star}}{M_{\odot}}\right)^{1.82}
    \left(\frac{40\,\mathrm{km\,s^{-1}}}{V_{\star}}\right)^{1.82}
    \left(\frac{2\times10^{4}\,\mathrm{AU}}{D_{\odot}}\right)^{0.91},
\end{equation}
for $\left(\frac{M_{\star}}{M_{\odot}}\right)\left(\frac{40\,\mathrm{km\,s^{-1}}}{V_{\star}}\right) < 0.6$, and
\begin{equation}
    N_{\star} = 12.83
    \left(\frac{M_{\star}}{M_{\odot}}\right)^{0.89}
    \left(\frac{40\,\mathrm{km\,s^{-1}}}{V_{\star}}\right)^{0.89}
    \left(\frac{2\times10^{4}\,\mathrm{AU}}{D_{\odot}}\right)^{1.78},
\end{equation}
otherwise. These expressions demonstrate that the number of comets dynamically injected into the planetary region rises steeply with decreasing encounter distance and lower stellar velocities, in qualitative agreement with the Hills scaling relation. 

Random stellar encounters not only lead to isotropy in the velocity distribution of Oort Cloud comets but also contribute to a slow diffusion of orbital energies that thermalizes the comets and replenishes the loss cone. The equilibrium achieved through this mechanism produces a steady flux of long-period comets toward the inner Solar System on a million-year time-averaged basis. Understanding the signatures of these encounters in the most eccentric Oort Cloud comets is essential for linking the dynamical perturbations induced by stellar flybys to the observed comet-shower rates within Jupiter’s orbit and their record in Earth’s geological and cratering history. 

\subsection{Rate of Comet Showers}

To quantify the expected enhancement in cometary flux following a stellar encounter, an analytical estimate of the instantaneous rate of comets penetrating the planetary region was derived. The approach is based on computing the fraction of an orbital period during which a comet resides within Jupiter’s orbital radius $R_{J}$ and comparing this quantity before and after the encounter. This fraction, multiplied by the total number of comets characterized by a given $(a,e)$ distribution, yields the total number of comets simultaneously present within the inner Solar System.

For a comet of semi-major axis $a$ and eccentricity $e$, the fraction of the orbital period spent at heliocentric distances $r<R_{J}$ can be obtained from Kepler’s second law, $dA/dt=\pi a b / T$, where $A$ is the area swept by the radius vector, $b=a\sqrt{1-e^{2}}$ is the semi-minor axis, and $T$ is the orbital period. The relevant fraction is therefore the ratio of the area enclosed by the orbital arc within $r<R_{J}$ to the total area of the ellipse.

Using the polar form of the orbit, $r(\theta)=\frac{a(1-e^{2})}{1+e\cos\theta}$, and adopting the small-angle approximation valid for nearly parabolic comets ($\theta\ll 1$), this can be written as
\begin{equation}
    r(\theta) \simeq a(1-e^{2})\!\left[1+\frac{e}{2(1+e)}\theta^{2}\right].
\end{equation}
The area enclosed by the orbit between $\pm\Theta$, corresponding to the points where $r(\Theta)=R_{J}$, is
\begin{equation}
    A = \int_{-\Theta}^{\Theta}\!\!\int_{0}^{r(\theta)} r'\,dr'\,d\theta 
      = \frac{a^{2}(1-e^{2})^{2}}{2}
        \!\int_{-\Theta}^{\Theta}\!\!\left(1+\frac{e}{1+e}\theta^{2}\right)\!d\theta,
\end{equation}
which evaluates to
\begin{equation}
    A = a^{2}(1-e^{2})^{2}\Theta.
\end{equation}
Solving $r(\Theta)=R_{J}$ gives
\begin{equation}
    \Theta = \sqrt{2\!\left(1+\frac{1}{e}\right)
    \!\left(\frac{R_{J}}{a(1-e)}-1\right)}.
\end{equation}
Substituting into the expression for $A$ and normalizing by the total orbital area $\pi a^{2}\sqrt{1-e^{2}}$ yields the fractional time spent within $R_{J}$:
\begin{equation}
    \frac{\Delta t}{T}
      = \frac{\sqrt{2}}{\pi}(1+e)
        \sqrt{\frac{1-e}{e}}
        \sqrt{\frac{R_{J}}{a(1-e)}-1}.
\end{equation}
This expression gives the instantaneous probability that a comet with parameters $(a,e)$ resides inside Jupiter’s orbit. 

Applying this relation to the simulated ensembles and comparing it with the comet counts shows that the maximum enhancement in comet-shower flux occurs within the first $\sim3$~Myr after the encounter and gradually decays over $\sim10$~Myr, consistent with the timescale required for perturbed comets to complete one or more orbital returns to the inner Solar System. This framework offers a quantitative means of interpreting the dynamical consequences of recent close-encounter candidates identified by \citet{Bailer_Jones_2022}, \citet{Bobylev_2022}, and \citet{Burgasser_2015}.

\begin{figure*}[ht!]
    \centering
    \includegraphics[width=\linewidth]{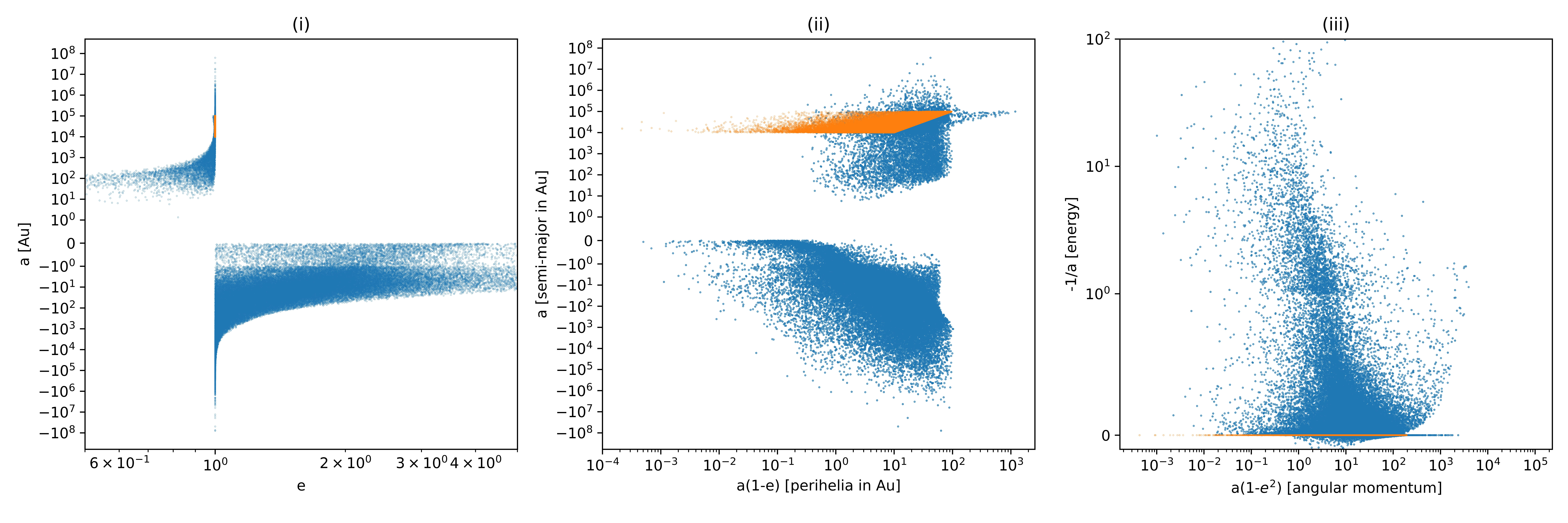}
    \includegraphics[width=\linewidth]{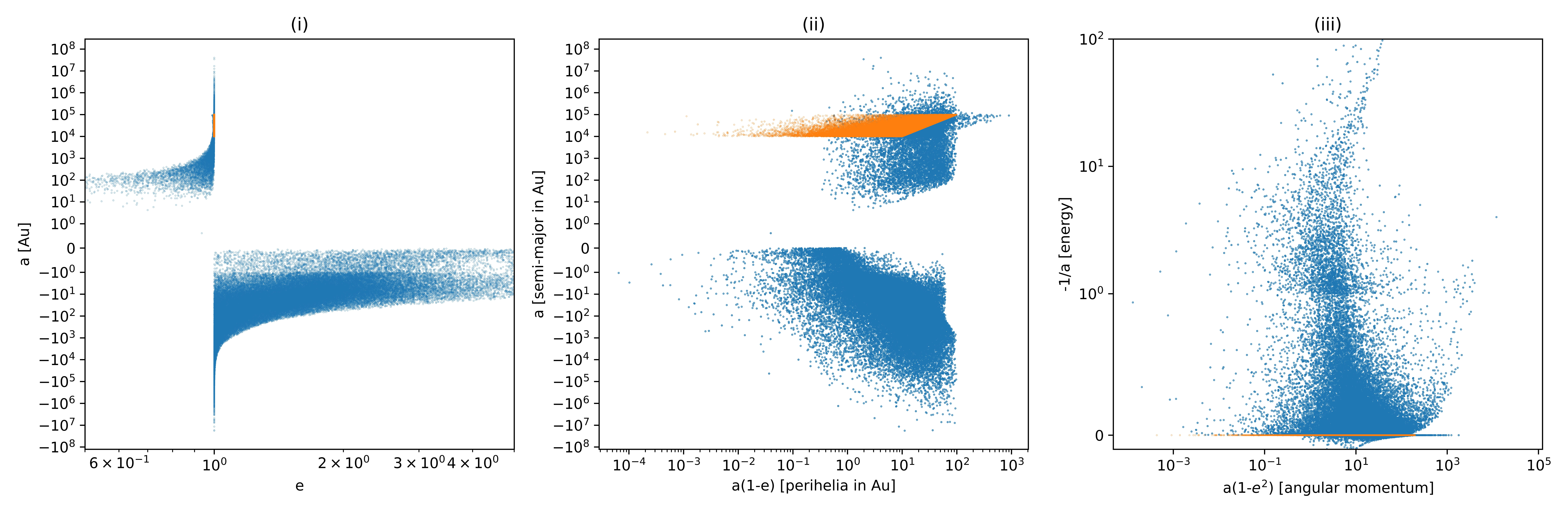}
    \includegraphics[width=\linewidth]{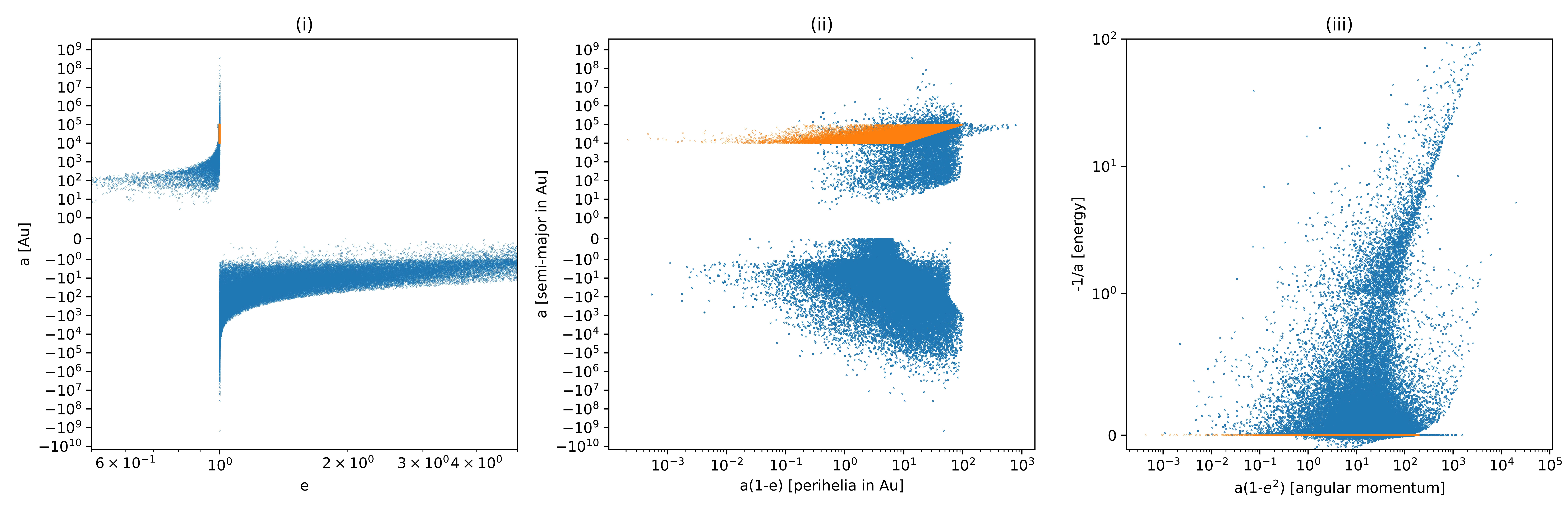}
    \caption{\textit{Note that the orange strip corresponds to the configuration before the encounter.} Scatter plots showing the dynamical response of comets before (orange) and after (blue) the encounter: 
    (i) semi-major axis vs.\ eccentricity; 
    (ii) semi-major axis vs.\ perihelion distance; 
    (iii) orbital energy vs.\ specific angular momentum; 
   The top, middle, and bottom set of plots correspond to encounter distances of $2\times10^{5}$, $2\times10^{4}$, and $2\times10^{3}$~AU, respectively.}
    \label{dynamics_all}
\end{figure*}

\subsection{Simulation}

To investigate the dynamical impact of an individual stellar flyby on the most eccentric comets in the OC, a direct $N$-body simulation was performed using the \textsc{Rebound} code \citep{Rein_2012}. The goal was to consider the imprint of the short-term impulse imparted by a passing solar-mass star on the most eccentric comets in the OC with the encounter parameters inferred from \textit{\textit{Gaia}}~DR2 \citep{Bailer_Jones_2018}. 

The simulation included the Sun, Jupiter, and 100,000 massless cometary test particles perturbed by a stellar flyby. The system was evolved using the \texttt{mercurius} hybrid symplectic integrator in heliocentric coordinates with time units of years, distance units of astronomical units, and mass units of solar masses. The integration time step was fixed at one year. The total integration duration was $10^{7}$~yr, corresponding to roughly full period evolution of the comets.

The initial comet population was designed to represent the high-eccentricity tail of a thermalized and isotropic Oort Cloud. One hundred thousand test particles were sampled from probability distributions proportional to $f(e)\propto 2e$ for the eccentricity and $f(a)\propto a^{-2}$ for the semi-major axis, consistent with a relaxed OC. The eccentricities were restricted to $0.999<e<1$, emphasizing the subset of comets that contribute most efficiently to inner Solar System flux. The semi-major axes were drawn from $10^{4}\,\mathrm{AU}<a<10^{5}\,\mathrm{AU}$. Orbital angles---inclination $i$, argument of perihelion $\omega$, longitude of ascending node $\Omega$, and true anomaly $f$---were uniformly distributed to produce an isotropic ensemble. 

The perturber was modeled as a one-solar-mass star on a rectilinear trajectory with perihelion distances of $0.01$, $0.1$, and $1$~pc and a velocity at perihelion of $V_{\star}=50\,\mathrm{km\,s^{-1}}$ (equivalent to $10.56\,\mathrm{AU\,yr^{-1}}$). The star was initialized at a distance of approximately $2\times10^{5}$~AU (near 1~pc) and assigned a constant approach velocity along the $y$-axis. The Sun was placed at the origin, and Jupiter, with $m_J=10^{-3}\,M_\odot$, $a_J=5.2$~AU, and $e_J=0.05$, was included to account for the dominant planetary contribution to hyperbolic comet ejection.

Figure~\ref{peri_se} displays the pre- and post-encounter perihelion distributions for three representative encounters at distances of $2\times10^{5}$, $2\times10^{4}$, and $2\times10^{3}$~AU, respectively. In each case, a noticeable population of comets migrates toward smaller perihelia ($q<5$~AU) after the stellar passage, with the exception of when the flyby penetrates the inner radius of the OC, corresponding to potential observable long-period comets entering the planetary region. As estimated by \citet{Emelyanenko2007}, the flux of new comets from the Oort Cloud into perihelia \(q<5\,\rm AU\) is approximately 10 per year.  This number agrees with the results of the numerical simulation in this work within an order of magnitude. The discrepancy might be due to overestimation of the number of comets in the OC, or the assumptions about full relaxation of the comets which is a key starting point of the numerical set up in this analysis. After the encounter in the simulation the number of comets entering within the Jupiter's orbit increase by a factor of $\sim 2$. Summed up over a million year period for all the twenty encounters this would amount to $\sim 40$. 

Although these numbers do not strictly scale with the analytic expectation $N\propto D_{\odot}^{-4}$, the behavior is consistent with the geometry of an isotropic, highly eccentric population. Most of the selected comets reside near aphelion during the stellar passage and are weakly affected by small variations in encounter distance, leading to a plateau in injection efficiency once the perturber penetrates within the Oort Cloud boundary. Closer encounters that pass through the inner OC tend instead to reduce the number of comets on Sun-penetrating trajectories by removing them entirely from bound orbits. This subtle interplay between geometric configuration and encounter strength explains the apparent departure from the simple analytic scaling and highlights the importance of modeling the phase-space distribution self-consistently.

The subsequent dynamical evolution of the comet ensemble is illustrated in Figure~\ref{dynamics_all}. Before the encounter, the population occupies a narrow locus in the semi-major-axis–eccentricity plane corresponding to $0.999<e<1$. After the encounter, the distribution broadens along curves of constant specific angular momentum, $l^{2}=GM_{\odot}a(1-e^{2})$, consistent with conservation of angular momentum during the impulse. Comets scattered toward smaller $a(1-e)$ values acquire perihelia within Jupiter’s orbit and form the source population for observable comet showers. The vertical component of the angular momentum ($l_{z}$) remains approximately conserved demonstrating that the perturbation acts primarily within the plane of the stellar passage. The observed dispersion reflects the small torque component exerted perpendicular to this plane. 

Overall, the simulation indicates that the perturbations caused by stellar flybys on the most eccentric comets in the OC do not have a substantial effect on the comet showers within the Jupiter's orbit, but play a major role in thermalization of these comets and making them unbound, transforming their elliptical orbits into hyperbolic ones. The approach used here isolates the instantaneous dynamical response, providing a high-resolution framework for quantifying the statistical enhancement of long-period comet fluxes in future Monte Carlo ensembles of encounters.

\begin{table}[h!]
    \centering
\caption{Phase-space parameters of the periodic companion candidate from \textit{Gaia} DR3.}
\label{tab:Huma}
    \begin{tabular}{cc} 
    \hline
    \hline
  \textit{Gaia} DR3 ID&2452167910719793664\\ 
  \hline
          Ra&22.164619 \\ 
  Dec&-14.968106 \\ 
  Parallax&74.896114	\\ 
  PmRa&-2.702113 \\ 
  PmDec&-28.048613 \\ 
  Radial Velocity&15.943527 \\ 
  \hline
 $\delta$ Ra& 0.165170\\ 
 $\delta$ Dec& 0.095928\\ 
 $\delta$ Parallax& 0.233388\\ 
 $\delta$ PmRa& 0.223920\\ 
 $\delta$ PmDec& 0.083165\\ 
 $\delta$ Radial Velocity& 1.943320\\
 \hline
 \hline
 \end{tabular}

\end{table}

\begin{figure}
    \centering
    \includegraphics[width=1.1\columnwidth]{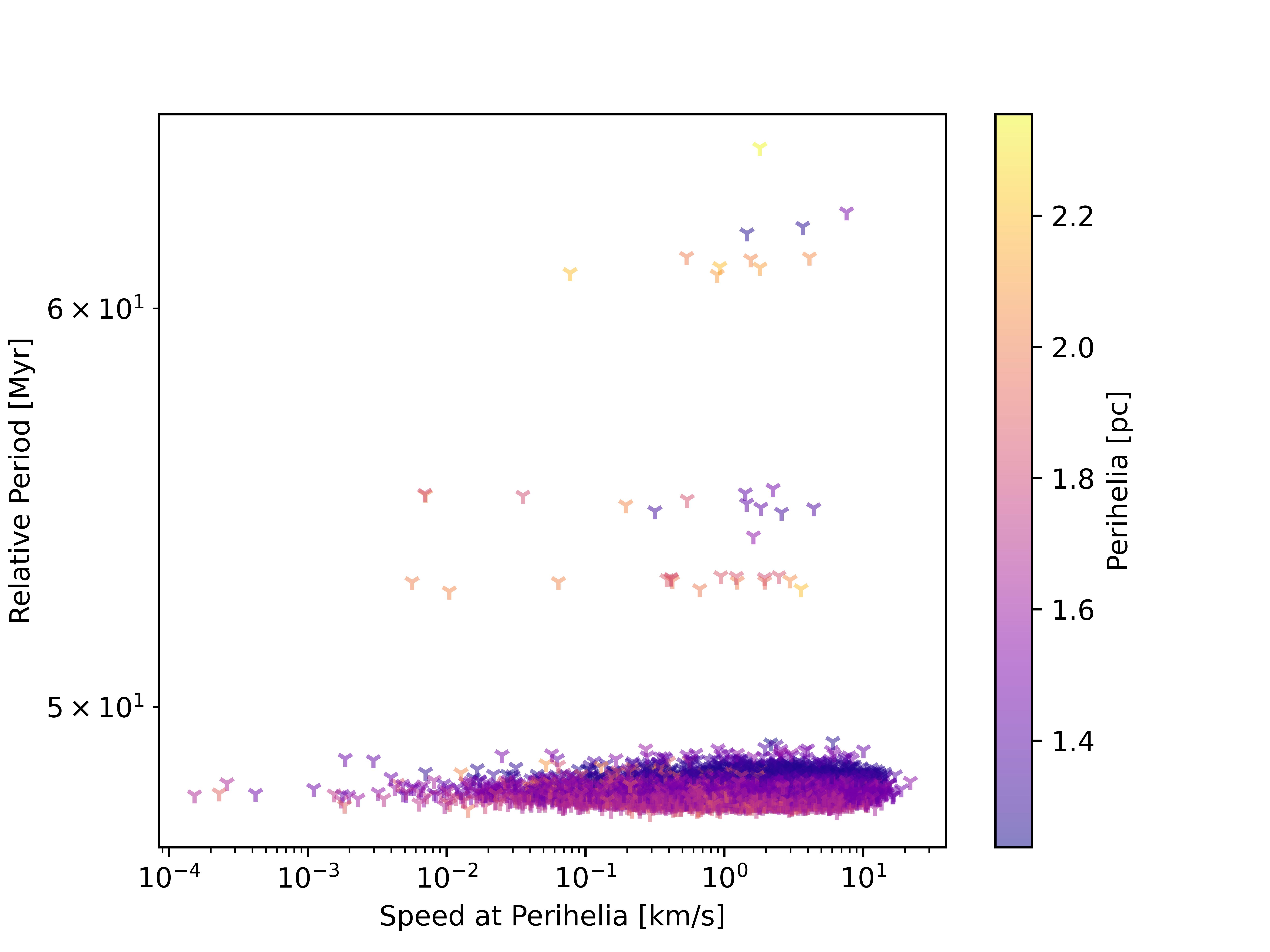}
    \caption{Relative period and speed at perihelia of the generated orbits for the companion candidate in \textit{Gaia} DR3,  color coded with their respective closest approach distance.}
    \label{fig:Huma}
\end{figure}

\bigskip

\section{Periodic Stellar Companion}

All previous searches for a gravitationally bound companion to the Sun have yielded null results. Nevertheless, the hypothesis remains of interest, both for its dynamical implications and for its potential relevance to the broader problem of identifying the Sun’s closest long-term stellar relatives. \textit{Gaia}~DR3 now provides the phase-space precision necessary to evaluate whether a star on a long-period, quasi-periodic orbit relative to the Sun might exist. As both bodies orbit the Galactic center and undergo their respective epicyclic motions, their relative trajectories can generate approximately periodic patterns. A companion approaching the Sun with a sufficiently small perihelion distance and a modest relative velocity could induce significant perturbations in the OC. The following subsections describe the time-dependent Galactic potential adopted in this work, the identification of a viable periodic companion candidate in \textit{Gaia}~DR3, and the resulting dynamical effects on the outer Solar System as quantified through direct $N$-body integrations.

\subsection{Galactic Potential Model}

To integrate the orbits of stars and the Sun within a realistic Galactic environment, the axisymmetric component of the gravitational field was represented by \textsc{MWPotential2014} \citep{Bovy_2015}. This composite potential consists of three analytic terms that approximate the mass distribution of the Milky Way:

\begin{enumerate}
    \item a power-law spherical bulge with an exponential cut-off,
    \item a Miyamoto--Nagai disk, and
    \item a Navarro--Frenk--White (NFW) dark matter halo.
\end{enumerate}

The bulge potential is expressed as
\begin{equation}
    \Phi_{\mathrm{bulge}}(r)
      = \Phi_{0}\!\left(\frac{r_{0}}{r}\right)^{\!\alpha}
        \exp\!\left[-\!\left(\frac{r}{r_{c}}\right)^{2}\right],
\end{equation}
where $\alpha=1.8$ is the inner power-law slope, $r_{0}=8$~kpc is the reference radius, and $r_{c}=0.24$~kpc is the exponential cut-off. This term represents the contribution of the Galactic bulge to the overall potential.

The disk component follows the Miyamoto--Nagai form,
\begin{equation}
    \Phi_{\mathrm{disk}}(R,z)
      = -\frac{G M_{\mathrm{d}}}
        {\sqrt{R^{2} + \left(a + \sqrt{z^{2}+b^{2}}\right)^{2}}},
\end{equation}
where $(R,z)$ are cylindrical coordinates and the scale lengths are $a=3$~kpc and $b=0.28$~kpc. This potential reproduces the observed thin disk structure of the Milky Way.

The dark matter halo is modeled using the Navarro--Frenk--White profile,
\begin{equation}
    \Phi_{\mathrm{halo}}(r)
      = -\frac{G M_{\mathrm{h}}}{r}
        \ln\!\left(1+\frac{r}{r_{s}}\right),
\end{equation}
with a characteristic scale length $r_{s}=16$~kpc. The relative amplitudes of the bulge, disk, and halo components are $0.05$, $0.60$, and $0.35$, respectively, calibrated to reproduce a circular velocity of $\simeq220$~km~s$^{-1}$ at the Solar radius $R_{0}=8$~kpc. These three components together constitute the standard \textsc{MWPotential2014} axisymmetric model implemented in the \textsc{Galpy} framework.

Non-axisymmetric contributions, namely the Galactic bar and spiral arms, were incorporated separately and are described in Appendix~A. The potential does not include the central supermassive black hole, whose gravitational influence is negligible beyond $\sim0.1$~kpc for the orbits considered in this analysis.

\begin{figure*}
    \centering
    \includegraphics[width=\textwidth]{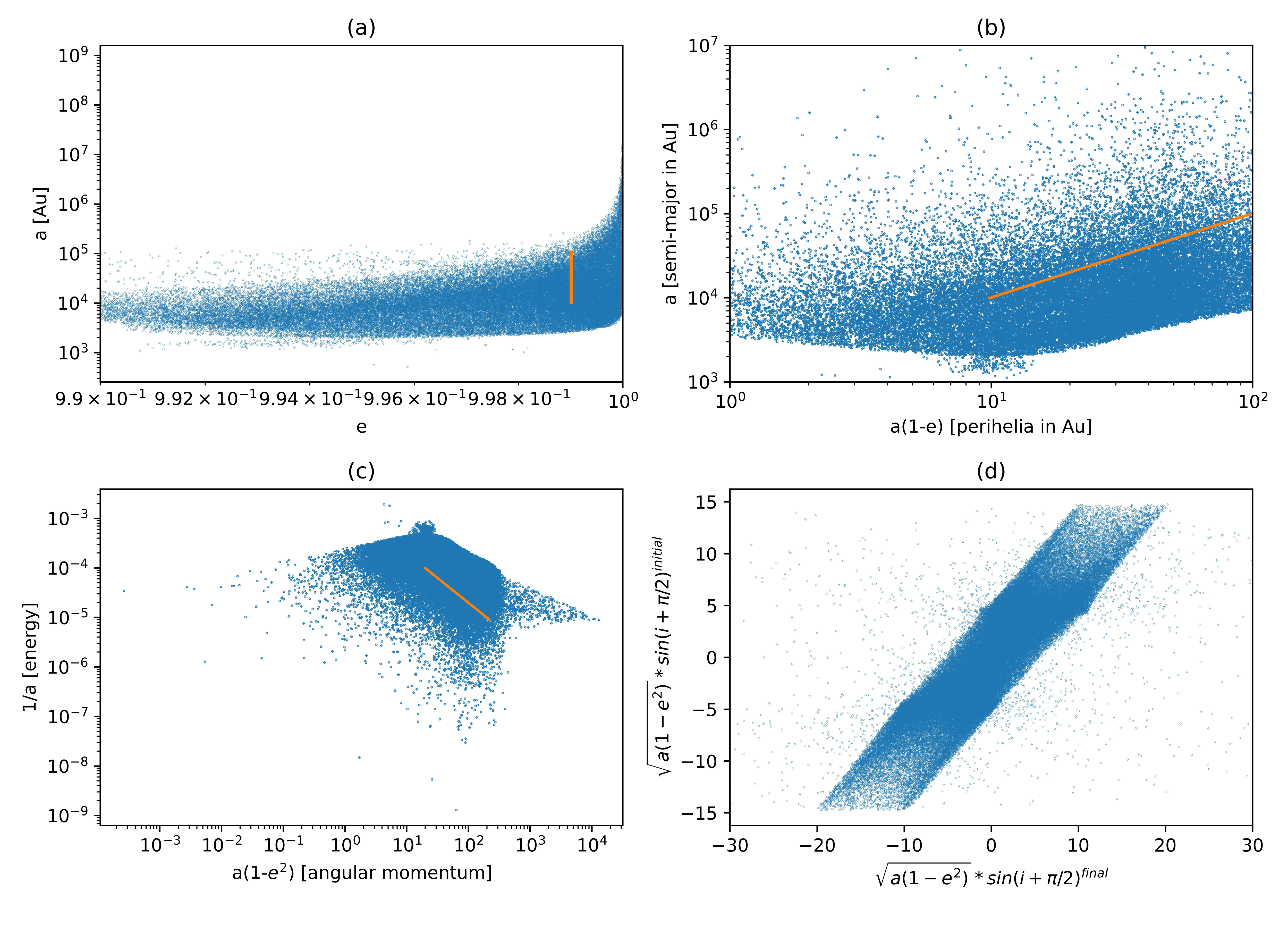}
    \caption{\textit{Note that the orange strip corresponds to the configuration before the encounter.} (a) The scatter in the semi-major axis v. eccentricity of the comets after the encounter. The initial distribution of the comets which is represented with the thin orange band gets scattered into more hyperbolic and more circular comets after the encounter. The region at the bottom right of the plot which corresponds to high $e$ and low $a$ is of interest for comet showers within the Jupiter's orbit. (b) Semi-major axis v. perihelia of the comets before and after the encounter. (c) Energy v. angular momentum of the comets before and after the encounter. (d) The z-component of the angular momentum before and after the encounter. The conservation of the z-component of the angular momentum is evident from the linear track of the figure, and the dispersion is due to the torque exerted by the encounter which is in the direction of the z-axis.}
    \label{dynamics_pc}
\end{figure*}

\subsection{Candidate Periodic Perturber from \textit{\textit{Gaia}} DR3}

A comprehensive orbital integration of all $\sim33$~million stars with complete six-dimensional phase-space information in \textit{\textit{Gaia}}~DR3 was performed in the Galactic potential described above. The trajectories were propagated for 500~Myr using \textsc{Galpy}, and their relative motions with respect to the Sun were analyzed. Among the full sample, a single star was found to exhibit quasi-periodic encounters with the Solar System, producing measurable perturbations in the modeled Oort Cloud population.

The fundamental parameters of this candidate are summarized in Table~\ref{tab:Huma}. The star’s perihelion distance relative to the Sun is $1$~pc, and its velocity at perihelion is approximately $1$~km~s$^{-1}$. Although no effective temperature is available in the \textit{\textit{Gaia}} dataset, a fiducial stellar mass of $0.6\,M_{\odot}$ was adopted. This value is consistent with the lower mass limit inferred from the empirical mass--temperature relation for main-sequence stars and with the minimum temperature ($\sim3100$~K) observed in the \textit{\textit{Gaia}}~DR3 sample. The adopted mass has only a minor effect on the perturbation amplitude, as the dominant parameters governing the encounter strength are the perihelion distance and the relative velocity.

Integration of the star’s Galactic orbit indicates a recurrent perihelion passage approximately every $45$~Myr, producing periodic gravitational impulses on the outer Oort Cloud. To quantify the statistical uncertainty, 10,000 Monte Carlo realizations of the star’s orbit were generated by sampling its six-dimensional Gaussian error distribution from \textit{\textit{Gaia}}~DR3. The resulting distribution of orbital periods and velocities at perihelion is shown in Figure~\ref{fig:Huma}.

The dynamical effects of this single encounter are illustrated in Figure~\ref{dynamics_pc}. As in the previous section, the pre-encounter population forms a narrow locus in the semi-major-axis--eccentricity plane corresponding to $0.999<e<1$. After the encounter, comets populate curves of constant specific angular momentum $l^{2}=GM_{\odot}a(1-e^{2})$, demonstrating approximate angular-momentum conservation during the impulse. The range of eccentricities broadens from $e_{\min}=0.999$ to $e_{\min}=0.83$, noting that most comets turn into hyperbolic orbits, while the corresponding spread in perihelion distances produces an increase in the number of comets entering the region $q<5$~AU. The conservation of the vertical component of angular momentum is evident in the Figure~\ref{dynamics_pc}(d), with the residual dispersion attributable to the weak vertical torque of the encounter.

Relative to the random-encounter simulations, all dynamical effects are amplified due to the star’s lower encounter velocity. Using the analytical expression derived in in previous sections, the number of comets with $q<R_{J}$ (where $R_{J}=5.2$~AU) increases by approximately a factor of $\sim 10$ compared to the unperturbed state. The resulting comet-shower events are expected to peak within $\sim0.3$~Myr of each perihelion passage and to persist for several Myr as the injected comets complete their inward orbits. Such periodic enhancements, repeating every $\sim50$~Myr, could in principle contribute to episodic increases in the terrestrial impact rate, a possibility that may be testable against the geological crater record. It would be important to conduct further analysis on the dependency of our results on the particular non-axis symmetric potential that has been utilized for the numerical simulation, and to follow up this search with stellar spectroscopy so its chemical abundance could be compared to that of the Sun. 

\section{Conclusion}

This work investigated two dynamical mechanisms capable of reshaping the most eccentric members of the OC: random stellar encounters and the coherent perturbations that may arise from a long-period stellar companion. Although distinct in origin, both processes act most efficiently on comets with large semi-major axes and nearly parabolic eccentricities, the subset responsible for delivering long-period comet showers into the planetary region. By combining analytical estimates with direct $N$-body simulations, we have linked Galactic-scale stellar dynamics to the observable comet flux in the inner Solar System.

Simulations of single solar-mass flybys with parameters characteristic of the \textit{\textit{Gaia}}-derived encounter distribution show that the most likely perturbations can temporarily increase the flux of comets with perihelia inside Jupiter’s orbit by a factor of order two. These enhancements persist for several Myr before the ensemble re-thermalizes through subsequent weak encounters and the Galactic tides in the long-run. The cumulative frequency of individual passages maintains the OC in a steady-state energy distribution and continuously replenishes the loss cone. This stochastic background sets the baseline level of long-period comet injection expected over the Sun’s history, consistent with earlier theoretical expectations regarding velocity-space diffusion and the long-term thermal equilibrium of the cloud.

In contrast, a long-term or quasi-periodic stellar companion can produce coherent perturbations that recur on fixed timescales and repeatedly target the same region of phase space. A full orbital integration of all $\sim 33$~million stars with complete phase-space information in \textit{\textit{Gaia}}~DR3 identified one candidate whose relative motion with respect to the Sun exhibits recurrent perihelion passages on $\sim 45$~Myr timescales. Monte Carlo sampling of its astrometric uncertainties confirms the stability of this periodicity. Simulated encounters of this type yield an order-of-magnitude increase in the number of comets entering the inner Solar System, with peaks occurring within a few $10^{5}$~yr of each passage and persisting over multi-Myr intervals. Such amplitudes are large enough to contribute to episodic structure in the terrestrial impact record, though this possibility requires more detailed modeling and comparison against both crater timescales and long-term Solar System resonant structure.

A periodic perturber would also be a natural candidate for a dispersed solar sibling. Establishing such a connection demands a more comprehensive exploration of Galactic potential uncertainties and, crucially, high-resolution spectroscopy to compare the candidate’s elemental abundance pattern to that of the Sun. Such observations could help determine whether the candidate traces back to the Sun’s birth environment, providing a valuable link between OC dynamics, solar sibling searches, and the early history of the Solar System. More broadly, the ability to identify stars with long-term dynamical affinity to the Sun opens the possibility of reconstructing past stellar associations that may have influenced the Solar System’s formation and early dynamical evolution.

The present analysis adopts several idealizations—including an isotropic and a simplified planetary architecture—that should be relaxed in future work. A larger Monte Carlo ensemble of encounters and the incorporation of the full planetary system will allow a more complete appraisal of the OC’s response to both stochastic and coherent perturbations. Upcoming \textit{\textit{Gaia}} data releases will further refine candidate companion trajectories, reduce astrometric uncertainties for stars exhibiting repeated approaches to the Sun, and may reveal additional dynamically interesting objects currently lost within the measurement noise.

Despite these limitations, the results presented here demonstrate that both random stellar flybys and possible long-period companions play central roles in shaping the dynamical landscape of the outer Solar System. Episodic surges in long-period comet activity arise naturally from these perturbations, and their signatures may persist in Earth’s geological and cratering record. Taken together, the findings point toward a view of the Oort Cloud not as a static reservoir but as a continuously evolving boundary layer between the Solar System and the broader Galactic environment, shaped by the cumulative imprint of its stellar neighbors. Future work combining dynamical modeling, chemo-dynamical tagging, and deeper \textit{\textit{Gaia}} astrometry may reveal whether the OC retains a detectable memory of both its random encounters and any long-period companion that may have accompanied the Sun over its Galactic lifetime.

\section{Acknowledgment}

I am deeply grateful to Phillip Nadjafov and Alexandria Lior Bader for their inspiring conversations, generous hospitality, and constant encouragement throughout this work. I thank Manoj Kaplinghat for originally suggesting this line of research during my undergraduate studies at the University of California, Irvine. I want to thank Steffani Grondin for the very helpful remarks and suggestions. 

This work made use of the \texttt{REBOUND} N-body code \citep{Rein2012} and the \texttt{galpy} Galactic dynamics library \citep{Bovy2015} for orbital and dynamical integrations. Computational resources were provided by the \textit{National Research Platform (NRP)}. We acknowledge the support and infrastructure enabling large-scale numerical simulations and data analysis. This work has made use of data from the European Space Agency (ESA) mission {\it \textit{Gaia}} (\url{https://www.cosmos.esa.int/\textit{Gaia}}), processed by the {\it \textit{Gaia}}
Data Processing and Analysis Consortium (DPAC, \url{https://www.cosmos.esa.int/web/\textit{Gaia}/dpac/consortium}). Funding for the DPAC
has been provided by national institutions, in particular the institutions
participating in the {\it \textit{Gaia}} Multilateral Agreement. The author acknowledges the use of ChatGPT-4.0 and Grammarly to check for grammar and improve the language and readability of the manuscript. The author reviewed and edited all generated text and take full responsibility for the final content.

\nocite{*}
\bibliographystyle{aasjournal}
\bibliography{references}

\appendix
\section{The Spiral Arms}
For the central bar and spiral arms structure we use the models suggested by \cite{Monari_2019}, and \cite{Cox_2002} and the values provided by \cite{Monari_2019}. The bar  potential provided by \cite{Monari_2019} has the form $$\Phi_{b} (R,\phi, z, t) = A_{b}(t) \frac{v_{0}^2}{3} U(r) (\frac{R}{r})^2 cos \gamma_{b},$$where $r^2 = R^2 + z^2$ is the spherical radius, $R_{b}$ is the length of the bar, $R_{0}$ is the Galactocentric radius of the Sun, and $v_{0}$ is the circular velocity at $R_{0}$, $$\gamma_{b}(\phi, t) \equiv 2 (\phi - \phi_{b} - \Omega_{b}t),$$ and $$U(r) \equiv \begin{cases}(\frac{r}{R_{b}})^{-3},& \text{for } r \ge R_{b},\\ (\frac{r}{R_{b}})^3 - 2,& \text{for } r < R_{b}. \end{cases}$$ $A_{b}(t) = A_{f} (\frac{3}{16} \xi^5 - \frac{5}{8} \xi^3 +\frac{15}{16} \xi + \frac{1}{2}), \; \text{given that }\xi = 2 \frac{t/T_{b} - t_{form}}{T_{steady}} - 1$, if $t_{form} \le \frac{t}{T_{b}} \le t_{form}+T_{steady}$, and $$A_{b}(t) = \begin{cases} 0,& \text{for } \frac{t}{T_{b}} < t_{form},\\ A_{f},& \text{for } \frac{t}{T_{b}} > t_{from}+T_{steady}, \end{cases}$$where $T_{b} = \frac{2\pi}{\Omega_{b}}.$ Following \cite{Monari_2019}, we choose for the simulations $\alpha = 0.01$, $\Omega_{b} = 52.2$ km s$^{-1}$ kpc$^{-1}$ so that $\Omega_{b} / \Omega(R_{0}) = 1.89$, $\phi _{b} + \Omega _{b} t_{e} = 25^{\circ}$, where $t_{e}$ corresponds to the present time. We also choose $t_{form} = -3 \: $Gyr and $T_{steady} = 3 \: $Gyr.

For the spiral arms we use the model suggested by \cite{Cox_2002} using the values given by \cite{Monari_2019}. Assuming that there are two main spiral arms patterns (namely, the Scutum-Centaurus and Perseus arms), then the contribution of the spiral arms to the total potential would be $$\Phi_{s}(R, \phi, z, t) = -\frac{A}{R_{s}KD} \cos \gamma_{s} [\text{ sech } (\frac{K z}{\beta} )]^{\beta},$$ where $$K(R) = \frac{2}{R \sin p},\;\beta (R) = K(R) h_{s}[1+0.4K(R)h_{s},$$ $$D(R) = \frac{1 + K(R) h_{s} + 0.3[K(R) h_{s}]^2}{1 + 0.3 K(R) h_{s}},$$ $$\gamma_{s} (R, \phi, t) = 2 [\phi - \phi_{s} - \Omega _{s} t + \frac{\ln (R/R_{s})}{\tan p}].$$ Here, $p$ is the pitch angle, $A$ the amplitude of the spiral potential, $h_{s}$ controls the scale-height of the spiral, and $R_{s}$ is the reference radius for the angle of the spirals. Following \cite{Monari_2019} we choose $R_{s} = 1$ ,  $\Omega _{s} = 18.9$ km s$^{-1}$ kpc$^{-1}$, $\phi _{s} + \Omega _{s}t = -26 ^ {\circ}$, and $p = -9.9^{\circ}$, and $A = 341.8$ km$^{2}$ s$^{-2}$ .

\section{Other Periodic Companion Candidates}
\begin{figure}[h]
    \centering
    \begin{minipage}[b]{0.45\textwidth}
        \centering
        \includegraphics[width=\textwidth]{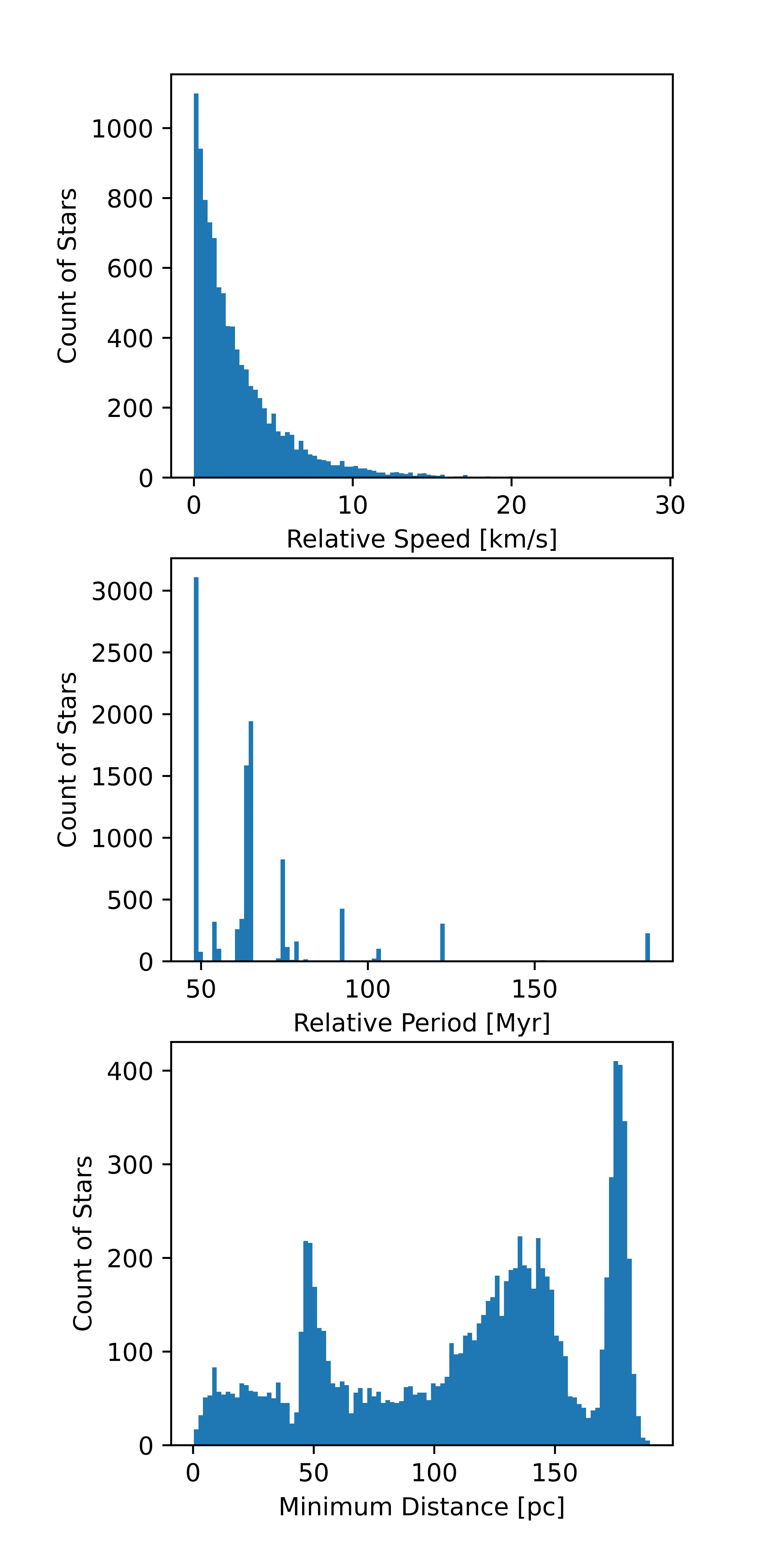} % Replace with your plot file
        \caption{Relative speed, relative period, and minimum distance distribution for the clone orbits of the star with \textit{Gaia} DR3 ID 2436007288814427264 }
        \label{fig:plot1}
    \end{minipage}
    \hspace{0.05\textwidth} % Adjust space between the plots
    \begin{minipage}[b]{0.45\textwidth}
        \centering
        \includegraphics[width=\textwidth]{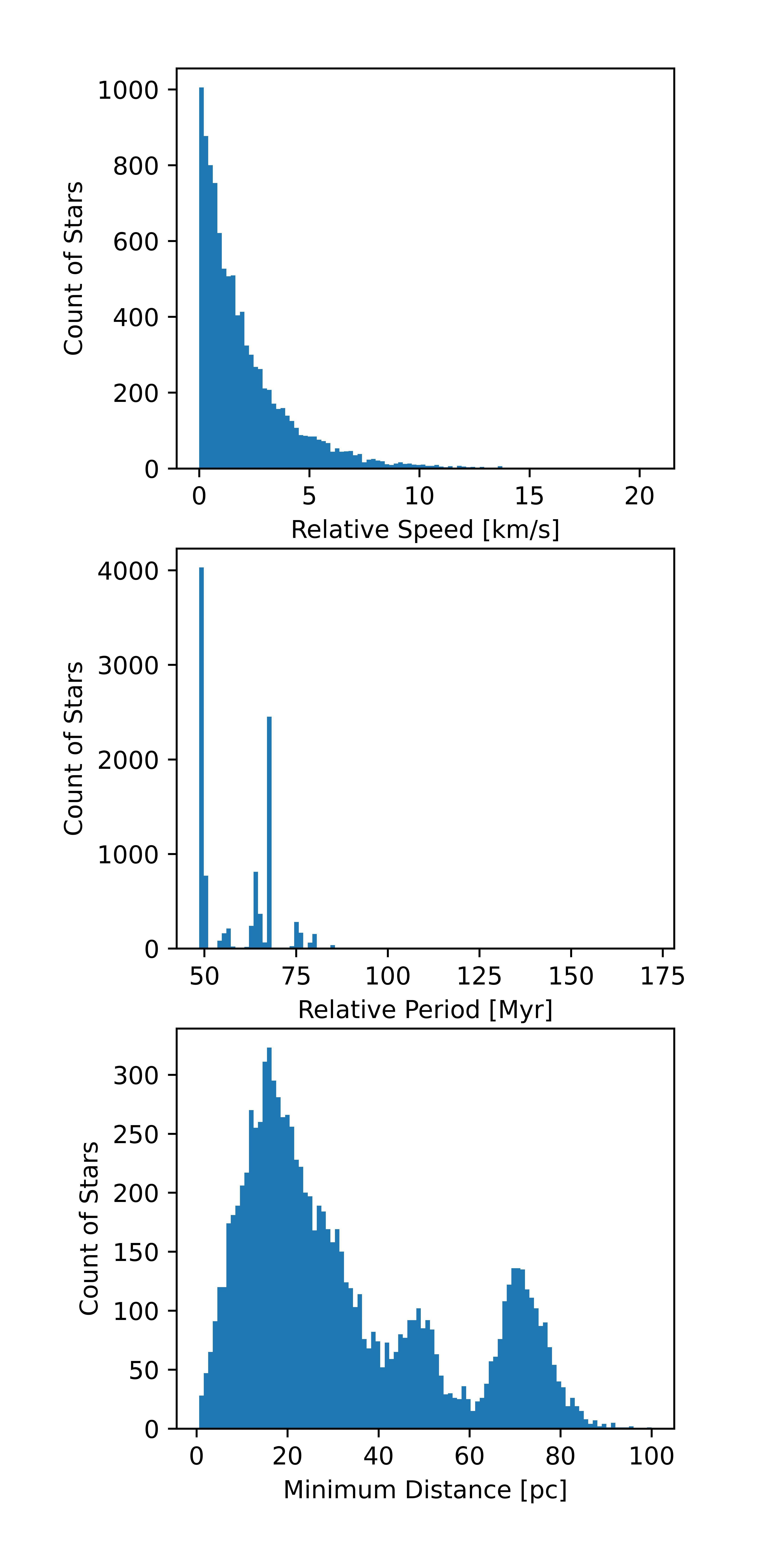} % Replace with your plot file
        \caption{Relative speed, relative period, and minimum distance distribution for the clone orbits of the star with \textit{Gaia} DR3 ID 1646520006920710400}
        \label{fig:plot2}
    \end{minipage}
\end{figure}

The phase-space parameters of the other two potential companion candidates are presented in Table \ref{appendixtable}. We run a simulation using 10,000 clone orbits within the error bounds, just as we did for the main companion, and the results are presented in Figure \ref{fig:plot1} and Figure \ref{fig:plot2}.
\begin{table}[t!]
    \centering
\caption{Phase-space parameters of the other periodic companion candidates from \textit{Gaia} DR3.}
\label{appendixtable}
    \begin{tabular}{ccc}
    \hline
    \hline
  \textit{Gaia} DR3 ID&      2436007288814427264&1646520006920710400\\ 
  \hline
          Ra& 
         
          357.389073 &22.164619\\ 
  Dec&      -8.566738 &-14.968106\\ 
  Parallax&      6.079200 &74.896114\\ 
  PmRa& 3.872023 & -2.702113 \\ 
  PmDec& -6.158308& -28.048613 \\ 
  Radial Velocity& 13.567369 & 15.943527 \\ 
  \hline
 $\delta$ Ra& 0.058997&0.165170\\ 
 $\delta$ Dec& 0.040789&0.095928\\ 
 $\delta$ Parallax& 0.068329&0.233388\\ 
 $\delta$ PmRa& 0.068033&0.223920\\ 
 $\delta$ PmDec& 0.046366&0.083165\\ 
 $\delta$ Radial Velocity& 2.668617&1.943320\\ 
 \hline
 \hline
 \end{tabular}

\end{table}
\end{document}